\documentclass[aps, prl, reprint, superscriptaddress]{revtex4-1}

\usepackage{graphicx}
\usepackage{siunitx}
\usepackage{textcomp}
\usepackage{xcolor}
\usepackage[normalem]{ulem}

\newcommand{\kp}{\ensuremath{k_\parallel}}
\newcommand{\El}{\ensuremath{E_0}}
\newcommand{\ltot}{\ensuremath{\lambda_{\text{tot}}}}
\newcommand{\lel}{\ensuremath{\lambda_{\text{el}}}}
\newcommand{\linel}{\ensuremath{\lambda_{\text{inel}}}}

\begin{document}

% Use the \preprint command to place your local institutional report
% number in the upper righthand corner of the title page in preprint mode.
% Multiple \preprint commands are allowed.
% Use the 'preprintnumbers' class option to override journal defaults
% to display numbers if necessary
%\preprint{}

\title{Non-universal Transverse Electron Mean Free Path through Few-layer Graphene}

\author{D.\ Geelen}
\affiliation{Huygens-Kamerlingh Onnes Laboratorium, Leiden Institute of Physics, Leiden University, Niels Bohrweg 2, P.O. Box 9504, NL-2300 RA Leiden, The Netherlands}

\author{J.\ Jobst}
\affiliation{Huygens-Kamerlingh Onnes Laboratorium, Leiden Institute of Physics, Leiden University, Niels Bohrweg 2, P.O. Box 9504, NL-2300 RA Leiden, The Netherlands}

\author{E.E.\ Krasovskii}
\affiliation{Departamento de F\'isica de Materiales, Universidad del Pais Vasco UPV/EHU, 20080 San Sebasti\'an/Donostia, Spain}
\affiliation{IKERBASQUE, Basque Foundation for Science, E-48013 Bilbao, Spain}
\affiliation{Donostia International Physics Center (DIPC), E-20018 San Sebasti\'an, Spain}

\author{S.J.\ van der Molen}
\affiliation{Huygens-Kamerlingh Onnes Laboratorium, Leiden Institute of Physics, Leiden University, Niels Bohrweg 2, P.O. Box 9504, NL-2300 RA Leiden, The Netherlands}

\author{R.M.\ Tromp}
\email{rtromp@us.ibm.com}
\affiliation{IBM T.J.Watson Research Center, 1101 Kitchawan Road, P.O.\ Box 218, Yorktown Heights, New York, New York 10598, USA}
\affiliation{Huygens-Kamerlingh Onnes Laboratorium, Leiden Institute of Physics, Leiden University, Niels Bohrweg 2, P.O. Box 9504, NL-2300 RA Leiden, The Netherlands}

\date{\today}

\begin{abstract}
In contrast to the in-plane transport electron mean-free path in graphene, the transverse mean-free path has received little attention and is often assumed to follow the \lq universal\rq\ mean-free path (MFP) curve broadly adopted in surface and interface science. Here we directly measure transverse electron scattering through graphene from 0 to 25\,eV above the vacuum level both in reflection using Low Energy Electron Microscopy and in transmission using electron-Volt Transmission Electron Microscopy. From this data, we obtain quantitative MFPs for both elastic and inelastic scattering. Even at the lowest energies, the total MFP is just a few graphene layers and the elastic MFP oscillates with graphene layer number, both refuting the \lq universal\rq\ curve. A full theoretical calculation taking the graphene band structure into consideration agrees well with experiment, while the key experimental results are reproduced even by a simple optical toy model.
\end{abstract}

% insert suggested PACS numbers in braces on next line
% insert suggested keywords - APS authors don't need to do this
%\keywords{}

%\maketitle must follow title, authors, abstract, \pacs, and \keywords
\maketitle

\section{Introduction}
The mean free path (MFP) of electrons, i.e., the average distance between scattering events, plays a key role in numerous areas of science and technology. As an electron moves through a medium (gaseous, liquid, solid, or plasma) it will undergo scattering which may be either elastic, or inelastic due to interaction with phonons, plasmons, nuclei or other electrons. The MFP of electrons determines many physical phenomena on all energy scales. At or near the Fermi level in a solid, the MFP is a key ingredient to the transport properties. For example, ballistic transport is only possible when the MFP is larger than the critical device dimension. At somewhat higher energies (several eV), where electrons can overcome the workfunction of a material and escape into the vacuum, the MFP determines from which depth below the surface an electron can escape. Thus, the probing depth of electrons in a Low Energy Electron Diffraction (or Microscopy) experiment, the electron escape depth in photoemission experiments, the efficacy of electron emission in electron sources and electron multipliers, and the spatial extent and resolution of electron interactions in Scanning Electron Microscopy, all depend on the electron MFP. At higher energies yet (1\,keV to 500\,keV), the MFP is of key concern in Transmission Electron Microscopy, and in plasmas for the interaction of energetic electrons with other plasma constituents and the solid surfaces in contact with the plasma. Finally, in the few MeV energy range, the relatively short MFP is useful in electron beam treatment of superficial cancers.

For electrons with vacuum energies from just a few eV to tens of keV, the MFP in solids is often assumed to be described by a \lq universal\rq\ curve, which implies the MFP to depend strongly on energy, but only weakly on material \cite{Seah1979}. This \lq universal\rq\ curve shows a minimum in MFP at energies of a few 10's of eV, increasing at both lower and higher energies. At the lowest energies not many excitation mechanisms other than phonons and intraband transitions are available for scattering, so the MFP is long, presumably up to a 100\,nm at 1\,eV according to Ref.\ \cite{Seah1979}. At somewhat higher energies (several eV to 10's of eV) surface and bulk plasmon excitations kick in, and the MFP drops to just $\sim$1\,nm. With further increasing energy, the cross-section for plasmon excitations decreases, while other excitations such as ionization have relatively small cross-sections, and the MFP again increases. Surface scientists use the MFP minimum to maximize surface sensitivity in electron probe and/or electron emission experiments \cite{Marsolais1991}.
This \lq universal curve\rq\ is widely accepted despite a dearth of experimental data below 30 eV. However, there is a more fundamental problem with the notion of a \lq universal\rq\ MFP curve \cite{Barrett2005-elastic, Naaman2007, Mullerova2012-method, Cazaux2012-reflectivity}: In a semi-infinite solid, an electron can only exist in a state with given energy and momentum if that state is allowed in the solid's electronic band structure. Consequently, electrons cannot propagate through a solid with an energy at which the bands are gapped.

Here we explicitly measure electron propagation through few layer graphene by recording the energy-dependent reflectivity and transmissivity in this low energy range directly. We demonstrate that the electron MFP strongly depends on electronic material properties in contrast to the \lq universal\rq\ curve. In particular, it changes greatly with the number of graphene layers and thus electronic interlayer resonances. This shows that, the motion of electrons through solids is intricately linked to band structure -- which is materials specific -- and thus, that electron mean free paths cannot be universal.

\begin{figure}[t]
	\centering  
	\includegraphics[width=\columnwidth]{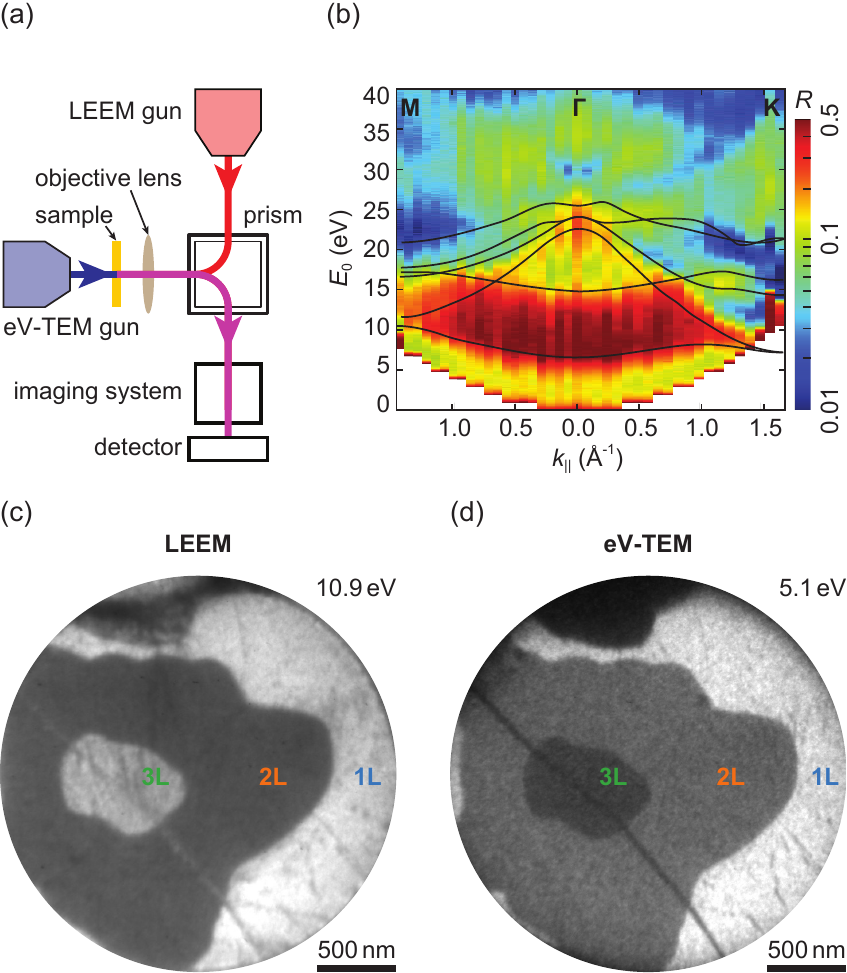}
	\caption{
		(a) Sketch of the ESCHER setup combining two electron guns for reflection (LEEM, red) and transmission (eV-TEM, blue) experiments. 
		(b) Angle-resolved reflected-electron spectroscopy of exfoliated bulk graphite showing electronic bands as minima and band gaps as maxima in electron reflection in agreement with band structure calculations (black lines). Reproduced from Ref.\ \cite{Jobst-ARRES-GonBN}. 
		(c) LEEM image of a free-standing membrane of 1, 2 and 3 layer graphene. 
		(d) The same area imaged in eV-TEM. Electron energy (indicated in to top right) in both images is chosen for optimal contrast.}\label{fig:TEM}
\end{figure}

%\section{Experimental}
We describe a set of experiments on thin graphene layers illuminated with electrons with kinetic energies \El\ in the range from 0 to 25\,eV where only the specularly reflected and the directly transmitted beams are present (first order Low Energy Electron Diffraction can only be excited above $\sim$28\,eV, which thus sets an upper limit to a straight-forward, quantitative interpretation). In addition to these energy-loss-free coherent beams, electrons scatter \lq thermally\rq\ (Debye-Waller scattering) and inelastically. Here, we quantitatively determine the elastic scattering fractions in both the specularly reflected and the transmitted electron beams as a function of energy, using the ESCHER aberration-corrected Low Energy Electron Microscope (LEEM) \cite{Tromp-AC1, Tromp-AC2, ESCHER} equipped with two distinct electron sources (see sketch in Fig. 1a). In a standard LEEM experiment, the sample is illuminated with an electron source from the front side of the sample (red in Fig. 1a), and an image is formed from reflected electrons. However, if the sample is sufficiently thin, one may also use an electron source located behind the sample (blue in Fig. 1a), and utilize electrons transmitted through the sample to form an image. Over the last few years we have developed such a capability enabling Transmission Electron Microscopy (TEM) experiments at electron energies of just a few eV \cite{Geelen-eV-TEM}, rather than 10's or 100's of keV as in conventional TEM. Using this \lq electron-Volt Transmission Electron Microscopy\rq\ (eV-TEM), we can thus study energy-dependent elastic transmission, in addition to energy-dependent elastic reflection using standard LEEM, on the same (thin) sample within the same instrument. 

To understand this energy dependence in an idealized system, let us assume that at a particular electron energy, at normal incidence ($\kp = 0$), our sample has a bandgap. In the absence of incoherent and inelastic channels, we would then expect a reflectivity of 1 and a transmissivity of 0 for electrons of that energy \cite{Slater1937-wave}. I.e., all electrons are elastically back-reflected as the sample electronic band structure has no states that would allow the electrons to propagate within the solid. One can \lq probe\rq\ the band structure of the solid above the vacuum level by measuring electron reflectivity as a function of energy and momentum \cite{Strocov1997, Jobst-ARRES, Wicki2016-mapping}. Figure 1(b) shows the results of such an Angle-Resolved Reflected Electron Spectroscopy (ARRES) experiment on bulk graphite reproduced from Ref.\ \cite{Jobst-ARRES-GonBN}. Reflectivity is plotted as function of energy and in-plane momentum \kp. The solid lines show theoretical band structure results from Ref.\ \cite{Hibino-thickness}, in good agreement with the data. Specifically, the high reflectivity (red) region across the Brillouin zone between $\sim$7 and 15\,eV corresponds with a large bandgap above the vacuum level. Conversely, regions of minimum reflectivity correspond to allowed states in the band structure that yield high transmission, and thus, can only be probed indirectly in these reflection experiments. Using eV-TEM, we can now investigate these high-transmission states directly, i.e.\ in transmission. In the following, we will focus on transmission and reflection data  for perpendicular incidence (i.e.\ at the $\Gamma$ point the center of the Brillouin zone) as a function of energy.
\begin{figure}[t]
	\centering  
		\includegraphics[width=\columnwidth]{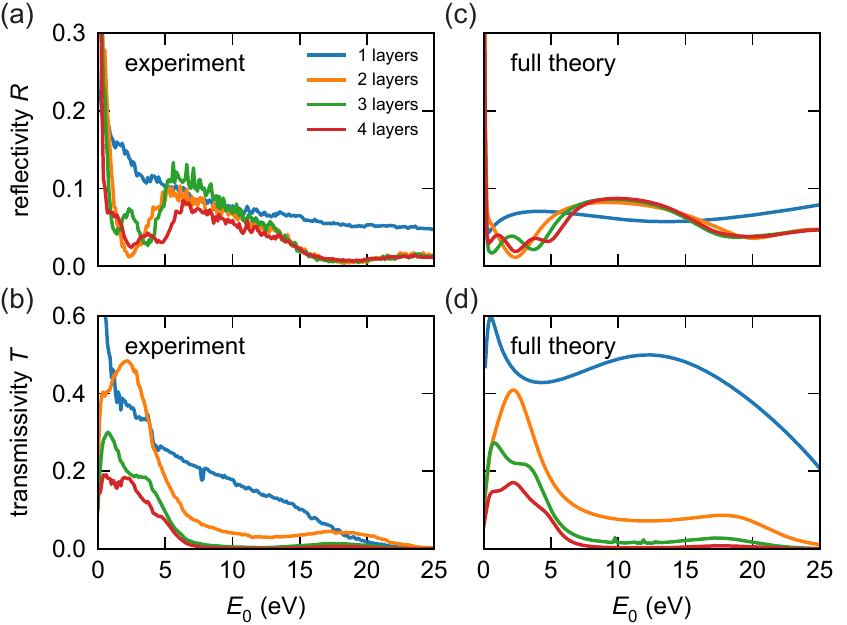}
	\caption{
		(a) Electron reflectivity $R(\El)$ as a function of landing energy $\El$ on 1--4 layer graphene. A general decreasing trend with strongly layer-dependent oscillations is observed. 
		(b) The electron transmissivity $T(\El)$ from the same areas also decreases with energy, but exhibits maxima where $R(\El)$ has minima. 
		(c,d) Theoretical predictions of $R(\El)$ and $T(\El)$ obtained by \emph{ab initio} methods \cite{krasovskii2004-augmented, nazarov2013-scattering-2D} reproduce the experimental data in all key features.}\label{fig:spectra}
\end{figure}

%\section{Results}
Figure 1(c) and (d) show LEEM ($\El=5.1$\,eV) and eV-TEM ($\El=10.9$\,eV) images, respectively. They are obtained on thin free-standing graphene of varying thickness of 1, 2 and 3 layers (in the following nLG will stand for n-layer graphene, with $n=1$--4). In LEEM, only the specularly reflected (0,0) LEED beam was used, and in eV-TEM only the directly transmitted beam. Inelastic electrons were removed from the signal by energy filtering, using the magnetic prism array as an efficient in-line energy filter \cite{Tromp-ARPES}. Thus, the image intensities in Fig.\ 1(c) and (d), normalized to the intensities of the incident electron beams, directly yield the elastic reflectivity $R$ and transmissivity $T$. Recording LEEM and eV-TEM images as a function of energy \El\ yields a laterally-resolved, spectroscopic data cube where reflectivity and transmissivity spectra can be extracted from every area. 
Figure 2 shows such $R(\El)$ LEEM (a) and $T(\El)$ eV-TEM (b) spectra for electrons with energies \El\ from 0--25\,eV, obtained on sample areas with 1--4 layers of graphene. 

In addition to a decrease with energy, strong modulations that depend not only on energy, but also on the number of graphene layers are visible for $R(\El)$ and $T(\El)$. For 2LG and thicker we find a broad maximum in reflectivity and minimum in transmission between 5 and 15\,eV, corresponding to the graphite bandgap seen in Fig.\ 1(b). For 1LG this feature is absent, indicating that this gap is a result of interlayer interactions. In fact, between 0 and 5\,eV we find $n-1$ reflection minima where $n$ is the number of graphene layers that are generally assumed to be caused by inter-layer transmission resonances that, eventually, merge into the broad minimum for many layers [see data on graphite in Fig.\ 1(b)] \cite{Hibino-thickness, feenstra2013-low}. 
Measuring corresponding maxima in transmissivity directly [Fig.\ 2(b)], we here confirm that reflection minima correlate with transmission maxima. Of course, for 1LG there is no interlayer scattering and thus no minimum/maximum. The energy dependence of reflectivity and transmissivity for all layer numbers is well reproduced by ab initio theory in Fig.\ 2(c) and (d), respectively. To obtain these, we calculate the ground-state potential of the n-layer graphene from first principles in the local density approximation and use it to obtain the scattering wave functions as described in Refs.\ \cite{krasovskii2004-augmented, nazarov2013-scattering-2D}. To account for inelastic effects, an energy-dependent optical potential is used in the scattering calculations. In addition, the theoretical reflectivity (obtained for a static lattice) is scaled down by a factor of 8 to fit the experimental reflectivity spectra which accounts for the enhanced Debye-Waller scattering in free-standing membranes.
Generally, both reflectivity and transmissivity (elastic signals) shown in Fig.\ 2(a, b) decrease with increasing energy, indicating increasing inelastic scattering. Using the quantitative $R(\El)$ and $T(\El)$ data, we can derive the inelastic MFP \linel\ and elastic MFP \lel, which combine to the total MFP \ltot. The total electron MFP \ltot\ can be obtained from the transmission data alone since both elastic and inelastic scattering give rise to a reduction of the elastically transmitted electron signal $I_{\text{et}}$. The elastic transmissivity $T$, shown in Fig.\ 2a, is thus given by
\begin{equation}\label{eq:ltot}
	T = \frac{I_{\text{et}}} {I_0} = \text{e}^{-\ltot / d}
\end{equation}
where $I_0$ the incident intensity and $d$ the sample thickness. Similarly, the inelastic MFP \linel\ can be obtained from the sum of reflected and transmitted intensities $I_{\text{er}}$ and $I_{\text{et}}$, as the total elastic signal is depleted by inelastic scattering only.
\begin{equation}\label{eq:linel}
	T + R = \frac{I_{\text{et}}} {I_0} + \frac{I_{\text{er}}} {I_0} = \text{e}^{-\linel / d}
\end{equation}
where $R$ is the elastic reflectivity shown in Fig.\ 2(a). Finally, the elastic MFP \lel\ is given by
\begin{equation}\label{eq:lel}
	\frac{1}{\ltot} = \frac{1}{\lel} + \frac{1}{\linel}~.	
\end{equation}
The experimentally measured \ltot, \linel\ and \lel\ are plotted in units of graphene layers as a function of electron energy in Fig.\ 3(a), (b) and (c), respectively.  
\begin{figure}[t]
	\centering  
		\includegraphics[width=\columnwidth]{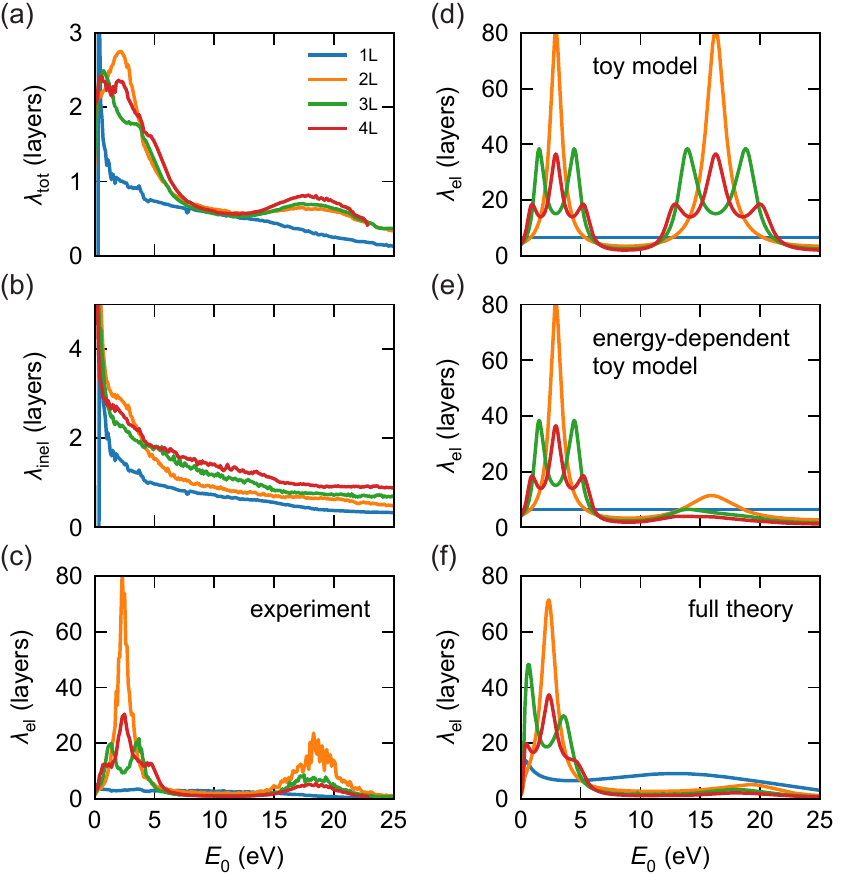}
	\caption{
		(a) Energy-dependent total MFP \ltot\ for electrons impinging on 1--4 layer graphene obtained from the spectra in Fig.\ 2 using Eq.\ (\ref{eq:ltot}). It decreases with energy, but is generally much lower than the \lq universal\rq\ curve predicts. 
		(b) The inelastic MFP \linel\ [Eq.\ (\ref{eq:linel})] shows a similar trend but the layer-dependent maxima are absent. 
		(c) The elastic MFP \lel\ [Eq.\ (\ref{eq:lel})] exhibits strong, layer-dependent peaks at the energies of high-transmission states where \lel\ is very long. The zoomed-in inset shows that the \lel\ is largest for monolayer graphene in the energy range from 6\,eV to 12\,eV.
		(d) An optical toy-model based on transfer matrices describes position and shape of the layer dependent oscillations in \lel\ well. 
		(e) When an energy-dependent absorption is taken into account, the experimental data in (c) is well described by this simple model over the full energy range. 
		(f) The \lel\ calculated from the \emph{ab initio} spectra in Fig.\ 2(c,d) also reproduces the experiment well over the full energy range.}\label{fig:MFP}
\end{figure}
Strikingly, the total MFP [Fig.\ 3(a)] is very short for all layer numbers, even at energies very close to 0\,eV. These values fall far short of the large numbers ($\sim$300 layers) suggested by the \lq universal\rq\ curve \cite{Seah1979}, even at energies below the graphene $\pi$-plasmon energy of $\sim$6\,eV \cite{Politano2014-plasmon}. Indeed, if we separate out the inelastic MFP [Fig.\ 3(b)] we find a monotonic decrease from $\linel \approx 3$ layers near 0\,eV to $\linel \approx 1$ layer at 25\,eV. While one may tentatively discern a somewhat steeper decrease in \linel\ at about 6\,eV, this is by no means a drastic effect, indicating the strong contribution of phonon and intra-band excitation losses at lower energies. 
The elastic MFP shown in Fig.\ 3(c) exhibits possibly the most interesting effects. Between 0 and 5\,eV, where the interlayer resonances occur, \lel\ is strongly modulated with the graphene layer number: one maximum for one interlayer resonance (2LG), two (three) maxima for 3LG (4LG). 
The coherent nature of this process calls into question the very name \lq mean free path\rq\ for \lel, but we keep using it for consistency with literature. 
Compared to the total and inelastic MFPs, \lel\ is quite large in this energy range with up to $\lel \approx 80$ layers for 2LG, and $\lel \approx 2$0--30 layers for 3LG and 4LG while it is smallest for 1LG. Conversely, between 6 and 15\,eV, \lel\ is largest for 1LG [see inset in Fig.\ \ref{fig:MFP}(c)] as the bandgap seen in thicker layers is not present for single layer, and electron tunneling may play a significant role. In the next band (15--22\,eV), an increased \lel\ is clearly visible while layer-number-dependent maxima are broadened due to inelastic effects, and no longer resolved. This second high transmission band was predicted by Feenstra \emph{et al}.\ \cite{feenstra2013-low, Feenstra2013-models}, but never before observed experimentally.

To arrive at a qualitative and intuitive understanding of these results, we turn to a simple toy model. In analogy with an optical multilayer, every graphene layer is modeled as a semitransparent boundary with reflectivity $R_1=|r|^2$ and transmissivity $T_1=|t|^2$ where $r$ and $t$ are the reflection and transmission amplitudes, respectively. The total reflectivity $R$ and transmissivity $T$ of a multilayer are then determined by the interference of all possible wave reflections and transmissions within the multilayer, while gaining a phase $\phi$ when propagating from layer to layer. Losses due to absorption, incoherent scattering, etc.\ are taken into account for every layer by setting $R_1 + T_1 < 1$. The energy-dependent $R(\El)$ and $T(\El)$ can be calculated using the transfer matrix approach (see Supplemental Materials \cite{SI} for details \nocite{Rayleigh1917-reflection, Gruneis2008-electron, Zhou2006-excitiations, Matsui2018-4pi}) frequently used in thin film optics e.g., to describe anti-reflective coatings \cite{Anders1967-thin-films, Zhan2013-transfer-matrix}.
Figures 3(d) shows \lel\ as a function of electron energy extracted from those $R(\El)$ and $T(El)$ using Eqs.\ (1--3) for the case with moderate absorption ($R_1 = 0.1$, $T_1 = 0.6$). The peaks corresponding to high-transmission states are clearly visible between 0 and 5\,eV and 15 and 20\,eV. Their position and dependence on layer number are in remarkable agreement with the experiment [Fig.\ 3(c)] given that other than the choice of $R_1$ and $T_1$, this is a parameter-free toy model (we use literature values of graphite for layer separation and work function, see Supplemental Material \cite{SI}). We obtain better agreement with the experiment at the second resonance between 15 and 20\,eV by taking losses increasing with energy into account [Fig.\ 3(e)]. We optimize $R_1(\El)$ and $T_1(\El)$ by comparing to LEEM specular reflectivity data of 1--8 graphene layers by Hibino \emph{et al}.\ \cite{Hibino2008-full-IVs} (see Supplemental Material \cite{SI} for details). We find that constant $R_1 = 0.1$, $T_1 = 0.6$ below the plasmon energy of 6\,eV where losses (phonons, intraband scattering, etc.) are not strongly energy-dependent, and then a smooth decrease to $R_1 = 0.033$, $T_1 = 0.2$ at 25\,eV to take account of $\pi$- and $(\sigma + \pi)$-plasmon losses above 6\,eV leads to excellent agreement with the data over the full energy range. This indicates that even this very simple optical toy-model captures the essential physics of electron scattering in multilayer graphene.

The full quantum mechanical approach [Fig.\ 3(f)], calculating the elastic MFP from the theoretical $R(\El)$ and $T(\El)$ shown in Fig.\ 2(c,d) again yields good agreement with the experimental data in Fig.\ 3(c). In this quantum mechanical picture, the maxima in \lel\ correspond to the transmission resonances. Comparison with the optical toy-model gives us the intuitive understanding that the enhanced transmission is the result of interlayer multi-reflection resonances of the electron waves. The energy range and scattering-induced broadening of the second high transmission band [15--20\,eV in Fig.\ 3(c)] is well described by both the toy model [Fig.\ 3(e)] and the full theory [Fig.\ 3(f)].

%\section{Discussion and conclusions}
We have presented the first direct measurements of elastic electron reflection and transmission data in the energy range between 0 and 25\,eV. While the \lq universal\rq\ MFP curve for electrons in this energy range would suggest very large mean free paths, up to 100\,nm at the lowest energies \cite{Seah1979}, we find that this prediction is far from true. Inelastic MFPs in single and multilayer graphene are just a few layers, even below the graphene $\pi$-plasmon energy of $\sim$6\,eV indicating that inelastic scattering due to phonon and intraband excitations plays an important role, on a par with plasmon excitation. Both \lel\ and \linel\ depend not only on electron energy, and on material (all carbon here), but also significantly on the details of the electronic structure above the vacuum level. The presence of interlayer resonances gives rise to high transmission, and very long \lel\ due to multilayer electron interference. This can explain the surprising fact that 1LG has the shortest total MFP over most of the energy range. These basic features are reproduced qualitatively by a simple toy model, and in detailed electronic structure + electron scattering theory. The high transmission/low reflection nLG graphene resonances correspond to electron anti-reflection coatings in our toy-model analogue. 
For other, more complex materials this simple toy model does not yield valid predictions. Already for other layered crystals such as hexagonal boron nitride \cite{Jobst-ARRES-GonBN} or transition metal dichalcogenides \cite{Krasovskii2007-band-mapping, DeJong2018-twist-angle} only the full \emph{ab initio} theory can describe the reflectivity spectra correctly. This indicates that the MFPs in these materials are also strongly affected by the band structure effects discussed here. Further measurements on those materials will, thus, elucidate our understanding of scattering of low-energy electrons with matter more broadly. 
Moreover, the observed transmission resonances strongly modify the so-called final state in Angle-Resolved Photo-Emission Spectroscopy (ARPES) for low photon energies \cite{Strocov2000-3D-unoccupied, Strocov2001-photoemission-graphite}. Together with the considerably shorter MFPs found here compared to the universal curve, this has broad implication on the interpretation of ARPES data at photon energies below 30\,eV.

The advent of eV-TEM in conjunction with LEEM, in a single instrument, has made it possible to study elastic reflection and transmission of low energy electrons from the same sample for the first time. This allows us to develop a more complete and detailed understanding of the interaction of low energy electrons with solids. These results challenge the perceived universality of electron MFP, and demonstrate that such universality cannot exist. The electronic structure of a material depends on its elemental composition, crystal structure, crystal orientation, and sample thickness and dominates scattering at low electron energies. 
The imaging and spectroscopic capabilities close to $\El = 0$ demonstrated here enables other eV-TEM experiments. For instance, using eV-TEM, we have succeeded in imaging single DNA origami molecules with electron energies below 5\,eV, where radiation damage appears to be negligible \cite{Germann2010-nondestructive, Longchamp2017-protein}, with strong contrast. Together with the projected spatial resolution below 2\,nm in an aberration-corrected instrument, eV-TEM promises new avenues for imaging and spectroscopy in physics, materials science, and life science.

\begin{acknowledgments}
We thank Marcel Hesselberth and Douwe Scholma for their indispensable technical support, Ruud van Egmond, Arthur Ellis, Raymond K\"ohler and Bert Crama for building the eV-TEM gun assembly and electronics, Aniket Thete for help with the transfer of graphene membranes and Martin van Exter for discussions on the transfer matrix approach.
This work was supported by the Netherlands Organisation for Scientific Research (NWO/OCW) via the VENI grant (680-47-447, J.J.) and the STW-HTSM grant (nr. 12789). It was supported by
the Spanish Ministry of Economy and Competitiveness MINECO, Grant No.\ FIS2016-76617-P.
\end{acknowledgments}

% If in two-column mode, this environment will change to single-column
% format so that long equations can be displayed. Use
% sparingly.
%\begin{widetext}
% put long equation here
%\end{widetext}

% Specify following sections are appendices. Use \appendix* if there
% only one appendix.
%\appendix*
%\section{Methods}
%\paragraph*{Sample fabrication} 

% Create the reference section using BibTeX:
\bibliography{ZZ_own-papers-2019_eV-TEM}

\end{document}